# Defining, Estimating and Using Credit Term Structures

# Part 2: Consistent Risk Measures

**Arthur M. Berd**
Lehman Brothers Inc.

**Roy Mashal**
Lehman Brothers Inc.

**Peili Wang**
Lehman Brothers Inc.

*In the second part of our series we suggest new definitions of credit bond duration and convexity that remain consistent across all levels of credit quality including deeply distressed bonds and introduce additional risk measures that are consistent with the survival-based valuation framework. We then show how to use these risk measures for the construction of market neutral portfolios.*

## INTRODUCTION

This paper continues our investigation of the consistent valuation methodology for credit-risky bonds (see Berd, Mashal, and Wang [2004a], cited hereafter as Part 1). In the previous article we have developed a set of term structures that are estimated using all the bonds of a given issuer (or sector) as a whole, rather than a specific bond of that issuer. In particular, our primary measure, the term structure of survival probabilities, clearly refers to the issuer and not to any particular bond issued by this issuer. However, when considering a particular bond, investors typically ask three questions:

- Is this bond rich or cheap compared with other bonds of the same issuer or sector?
- How much excess return does this bond provide for taking on the credit risk?
- How can we monetize these relative values, once we measure them?

The answer to the first question lies in the comparison of the observed bond price with the fair value price given by the fitted issuer credit term structures. The OAS-to-Fit measure, introduced in Part 1, gives an unambiguous and consistent answer to this question, free of biases associated with the term to maturity or level of coupon, which plague the conventional spread measures.

The answer to the second question then becomes straightforward, since we have already determined in the previous step the term structure of "fair value" par spreads of the issuer with respect to the underlying credit risk-free market. By adding the consistent issuer-specific and bond-specific spread measures we are able to give a robust definition of a bond spread and sidestep the ambiguities associated with non-par bond excess return estimation.

In order to answer the last question one must devise a recipe for hedging and risk managing credit bonds, which of course requires calculation of various sensitivity measures. Derivation of such measures and in particular the consistent definition of a bond's duration and convexity, as well as the bond's sensitivity to hazard rates and recovery values within the survival-based valuation framework are the objectives of the present paper.

We will show that the correctly defined duration of credit bonds is often significantly shorter than the widely used modified adjusted duration. This disparity helps explain the fact that high yield bonds do not have quite the same degree of interest rate sensitivity as high grade ones and is especially evident for distressed bonds, for which a 10-year maturity bond may have a duration as low as 1 year. This fact is well known to portfolio managers qualitatively – our paper provides its quantitative formulation.

The flip side of the same coin is the apparent negative correlation between interest rates and the conventionally defined credit spreads (OAS). It results in a similar effect of a dampened effective duration of credit bonds, as explained in Berd and Ranguelova (2003) and Berd and Silva (2004). We argue that a large portion of this negative correlation is "optical" in nature and is due to a misspecification of credit risk by the conventional OAS spread measures.

We also show that what is commonly regarded as a convexity measure for (both credit and Treasury) bonds is also a "duration" measure with respect to interest rate curve steepening/flattening moves. This is an important observation because the so-called "convexity trades" often under- or outperform not due to directional changes in interest rates, but because of the changes in the shape of the curve – as was the case, for example, during the past year and a half.

Finally, we present a concise and simple recipe for setting up well-hedged portfolios of bonds that generalizes the well-known duration-neutral and barbell trading strategies. We discuss how to choose the risk dimensions with respect to which one might wish to be hedged and how to find optimal security weights in the corresponding portfolios.

## SURVIVAL-BASED MODELING OF CREDIT-RISKY BONDS

Let us start with a brief reminder of the survival-based valuation methodology, following Part 1 of this series. Consider a credit-risky bond that pays fixed cash flows with specified frequency (usually annual or semi-annual). According to the fractional recovery of par assumption, the present value of such a bond is given by the *expected discounted* future cash flows, including the scenarios when it defaults and recovers a fraction of the face value and possibly of the accrued interest, discounted at the risk-free (base) rates.

By writing explicitly the scenarios of survival and default, we obtain the following pricing relationship (see Duffie and Singleton [2003] and Schonbucher [2003] for a detailed discussion of general pricing under both interest rate and credit risk):

[1]
$$\begin{aligned} PV(t) \;=\; & \sum_{t_i>t}^{t_N=T} \left( CF^{prin}(t_i) + CF^{\text{int}}(t_i) \right) \cdot \mathrm{E}_t \left\{ Z_{t_i} \cdot I_{\{t_i<\tau\}} \right\} \\ & + \int_t^T \mathrm{E}_t \left\{ \left( R_p \cdot F^{prin}(\tau) + R_c \cdot A^{\text{int}}(\tau) \right) \cdot Z_\tau \cdot I_{\{u<\tau\le u+du\}} \right\} \end{aligned}$$

The variable $\tau$ denotes the (random) default time, $I_{\{X\}}$ denotes an indicator function for a random event *X*, $Z_u$ is the (random) credit risk-free discount factor, and $\mathrm{E}_t\{\bullet\}$ denotes the expectation under the risk-neutral measure at time $t$.

The first sum corresponds to scenarios in which the bond survives until the corresponding payment dates without default. The total cash flow at each date is defined as the sum of principal $CF^{prin}(t_i)$, and interest $CF^{\text{int}}(t_i)$, payments. The integral corresponds to the recovery cashflows that result from a default event occurring in a small time interval $[u, u+du]$, with the bond recovering a fraction $R_p$ of the outstanding principal face value $F^{prin}(\tau)$ plus a (possibly different) fraction $R_c$ of the interest accrued $A^{\text{int}}(\tau)$.

Following the market convention, we assume in the following that recoveries happen discretely only on coupon payment dates, and that interest rates and recovery rates are independent of the default arrival. For the case of fixed-coupon bullet bonds with coupon frequency *f* (e.g. semi-annual *f=2*) and no recovery of the accrued coupon, this leads to a simplified pricing equation:

$$PV = Z_{base}(t_N)\cdot Q(t_N) + \frac{C}{f}\cdot\sum_{i=1}^{N} Z_{base}(t_i)\cdot Q(t_i) + R_p\cdot\sum_{i=1}^{N} Z_{base}(t_i)\cdot D(t_{i-1},t_i) \qquad [2]$$

We have dropped for simplicity the argument denoting the valuation time *t*. The probability $D(t_{i-1},t_i)$ that the default will occur within the time interval $[t_{i-1},t_i]$, conditional on surviving until the beginning of this interval, is related to the survival probability in a simple, reflecting the conservation of total probability:

$$D(t_{i-1},t_i) = Q(t_{i-1}) - Q(t_i) \qquad [3]$$

In Appendix A we derive a generic continuous-time approximation to the exact formula for the clean price of a fixed-coupon credit bond expressed through the term structure of the instantaneous forward interest rates and hazard (forward default) rates. This formula is quite accurate across all values of coupons and for all shapes and levels of the underlying interest rate and hazard rate curves. While such an approximation is superfluous for numerical computations, it comes in very handy for analytical estimates of bond risk measures, which is why we show it here and will use it in the next section.

## RISK MEASURES FOR CREDIT BONDS

In order to risk manage credit bond portfolios one must first calculate various sensitivity measures. Here we define a bond's duration and convexity, as well as its sensitivity to hazard rates and expected recovery values in a manner consistent with the survival-based valuation framework. As it turns out, the newly introduced risk measures are often substantially different from the commonly used ones such as modified adjusted duration, spread duration and convexity (see Fabozzi [2000] and Tuckman [2002] for standard definitions).

### Interest Rate Duration

There are many ways to define duration. We can think of duration as the sensitivity to changes in interest rates. We can also think of duration more generally as sensitivity to changes in non-credit related discount rates (such as the issue-specific OAS-to-Fit rate).

The two definitions become identical if we take the continuously compounded forward rates to be the primary variable with respect to which we measure the sensitivity. A parallel shift in instantaneous forward rates and an equal constant shift in OASF (the non-credit risk related pricing premium/discount) would cause an identical change in the bond's price. Had we defined the interest rate sensitivity with respect to some other measure of rates, such as a parallel shift in par yields, this equality would not hold.

The definition of duration in such terms is a direct modification of the well-known Macaulay duration. If we calculate the survival-based duration as the (minus) log-derivative of the bond's clean price with respect to OASF it becomes equal to the weighted time to cash flows, where the weights reflect not only the present value of the cash flows (as in the conventional Macaulay duration) *but also the probability of realization of the cash flow*:

$$
\begin{aligned}
D \;=\; & \left[ \sum_{i=1}^{N} t_i \cdot \frac{C}{f} \cdot Z_{base}(t_i) \cdot Q(t_i) \cdot e^{-OASF \cdot t_i} + t_N \cdot Z_{base}(t_N) \cdot Q(t_N) \cdot e^{-OASF \cdot t_N} \right. \\
& \left. + \sum_{i=1}^{N} t_i \cdot R_p \cdot Z_{base}(t_i) \cdot \left( Q(t_{i-1}) - Q(t_i) \right) \cdot e^{-OASF \cdot t_i} \right] \cdot \frac{1}{P}
\end{aligned} \qquad [4]
$$

The survival-based effective duration is always less than the classical Macaulay duration, reflecting the positive probability of receiving earlier (and larger) cash flows in the case of default. Depending on the level of the implied default rates, the differences can be quite large, as shown in Figure 1. It follows the changes in a particular Calpine bond (CPN 8.5 2/15/2011) as the company underwent different levels of distress during the past three years.

**Figure 1. Survival-based duration vs. conventional duration, CPN 8.5 2/15/11**

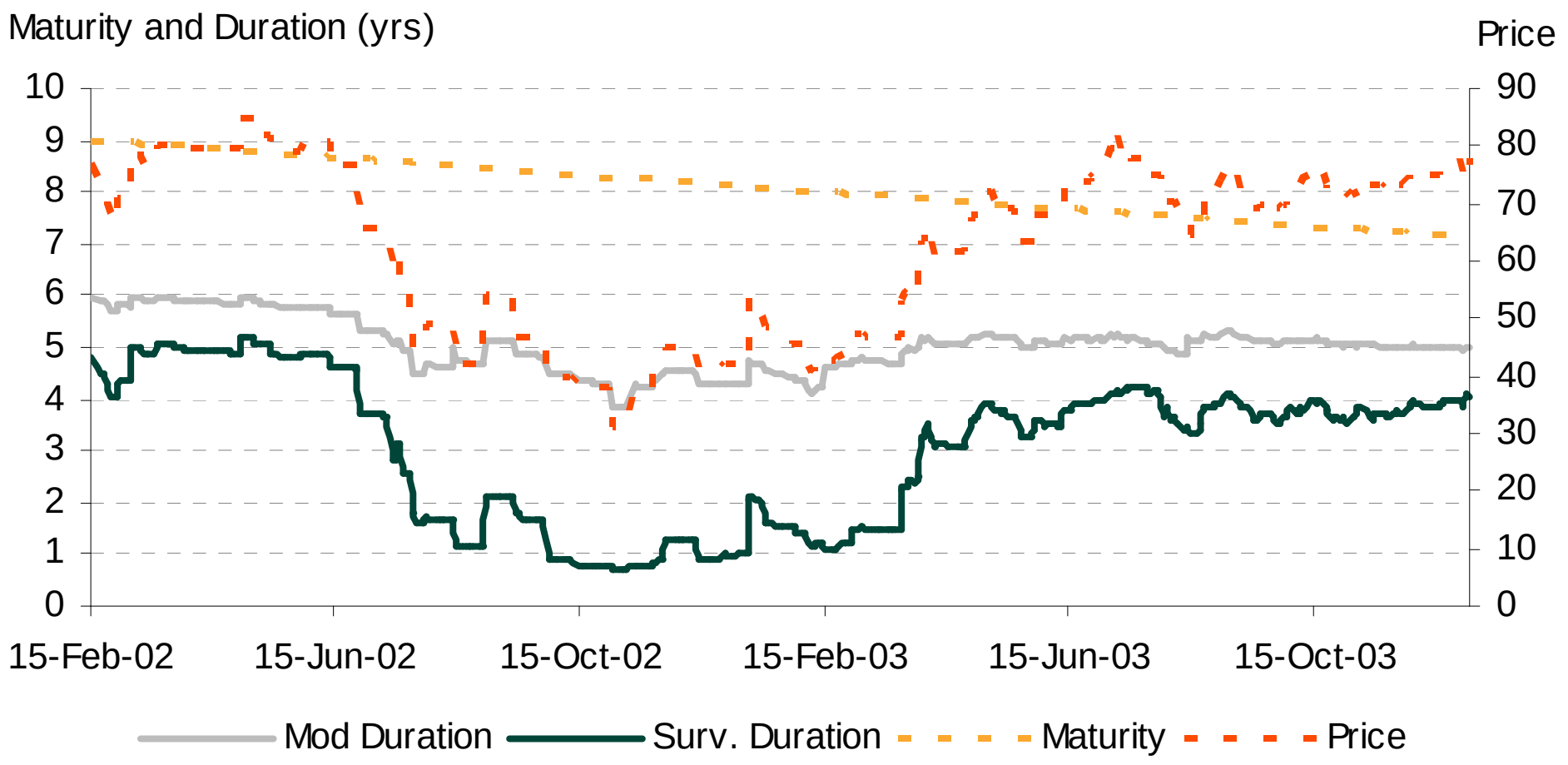

We see in particular that during 2002, as the company was deeply distressed and the bonds were trading at very low prices, the survival-based duration was as low as 1 year, compared with the conventional modified duration which decreased only slightly from the normal levels to around 4-year value. This is a very significant difference for a bond that still had more than eight years to maturity at the time. One could interpret it in terms of a much shorter "expected life" of the distressed security compared with its nominal maturity.

Many high yield portfolio managers are well aware of the propensity of distressed bonds to have much lower interest rate sensitivity than that prescribed by the conventional modified duration, but until recently few have been able to quantify this effect. Equation [4] gives a precise definition to this intuition, and Figure 1 demonstrates how important it can be.

## The Apparent Negative Rates-Spreads Correlation

As we just demonstrated, adoption of the survival-based valuation methodology leads to a significant decrease in the forecasted sensitivity to interest rates, even for relatively high grade (low credit risk) credit bonds. This effect should not be entirely surprising to credit portfolio managers. However, it has been usually discussed under the guise of a seemingly unrelated issue of the correlation between interest rates and spreads.

In two previous papers (Berd and Ranguelova [2003] and Berd and Silva [2004], see Appendix C for a brief recap) we have documented the empirical evidence for negative correlation between rates and spreads, and have explained how this results in a lower effective duration of credit bonds. In the framework of commonly used modified durations and OAS durations, the lower effective "duration" of credit bonds comes from the fact that (statistically) spreads tend to get tighter when rates get higher and vice versa. Thus, the movement of spreads tends to offset the movement of interest rates, and the total yield of credit bonds changes by only a fraction of the amount by which the Treasury (or LIBOR) rate changes. This, in turn, means that the expected price impact of a 1bp move in interest rates will be less than what one should expect by simply looking at a bond's conventional (modified adjusted) duration. The resulting "effective duration" multiplier is around 90% for investment grade bonds (Berd and Silva [2004]) and drops to 25% or less for high yield (Dynkin, Hyman, Konstantinovsky [2004]).

The net effect is consistent with that obtained from a fundamentally different model presented in a previous subsection. Why is this the case? Why do spreads, defined in a manner inconsistent with the survival-based valuation framework, conspire to move in such a way as to produce an effect that is similar to the correct model?

To answer this question, let us consider what would the survival-based valuation framework look like if we believed in the strippable discounted cash flows methodology that is the underpinning of OAS (see O'Kane and Sen [2004] for detailed definitions of conventional spread measures). For convenience, we will use the continuous-time approximation defined in Appendix A and ignore the small correction terms. We will also assume for simplicity flat interest rate, hazard rate and spread curves. On the left hand side of equation [5] we write the price of the bond with coupon *C* and maturity *T* using the valuation based on the conventional spread *S*, and on the right hand side the same price expressed under the survival-based methodology.

[5]
$$\frac{C}{r+S}\cdot\left(1-e^{-(r+S)\cdot T}\right)+e^{-(r+S)\cdot T}=\frac{C+h\cdot R_p}{r+h}\cdot\left(1-e^{-(r+h)\cdot T}\right)+e^{-(r+h)\cdot T}$$

This relationship may be considered as a parametric definition of the conventional spread *S* as a function of interest rate *r* and hazard rate *h*. This function can be easily obtained numerically by solving the equation for *S*, but it is more elucidating to see the relevant dependencies using the following approximate analytical solution, obtained by solving for the small correction in an expansion $S\approx h\cdot\left(1-R_p\right)+\varepsilon$ in the limit $\left(r+h\right)\cdot T<<1$:

[6]
$$S(r,h)\approx h\cdot\left(1-R_p\right)+\frac{1}{2}\cdot R_p\cdot h\cdot T\cdot\left(C-r-h\cdot\left(1-R_p\right)\right)$$

The correction to the "credit triangle" formula is proportional to the amount by which the coupon differs from the par level, times the total default probability – in other words the spread bias is related to the risk of losing the bond's price premium. We can see from this formula that if we keep the hazard rates constant, then rising interest rates lead to falling conventional bond spreads. There are of course other, less technical explanations, related to common economic driving factors, etc. But the demonstrated "optical" co-movement induced by the inherent biases of the OAS as a credit measure can account for a significant portion of the observed negative correlation between rates and spreads.

### Interest Rate Convexity and Twist Duration

Since the survival-based duration is significantly different from the modified duration, it is not surprising that the convexity measure will also be very different. The convexity is defined as the second derivative of the bond price with respect to yield change, expressed as a fraction of the bond's price. We remind the reader that in our definition, the derivative with respect to a parallel shift in forward interest rates is precisely equal to the derivative with respect to a change in OAS-to-Fit.

$$\Gamma = \frac{1}{P} \cdot \frac{\partial^2 P}{\partial r^2}$$

[7] $$= \left[ \sum_{i=1}^{N} t_i^2 \cdot \frac{C}{f} \cdot Z_{base}(t_i) \cdot Q(t_i) \cdot e^{-OASF \cdot t_i} + t_N^2 \cdot Z_{base}(t_N) \cdot Q(t_N) \cdot e^{-OASF \cdot t_N} + \sum_{i=1}^{N} t_i^2 \cdot R_p \cdot Z_{base}(t_i) \cdot \left(Q(t_{i-1}) - Q(t_i)\right) \cdot e^{-OASF \cdot t_i} \right] \cdot \frac{1}{P}$$

Survival-based convexity follows a similar pattern to the duration in terms of its deviation from the conventional measure, decreasing rapidly as the implied default rates rise.

We can use the following approximation to estimate the price impact of changes in the bond-specific non-credit related discount/premium encoded in the OAS-to-Fit measure:

[8] $$\frac{P(\Delta OASF)}{P_0} \approx 1 - D \cdot \Delta OASF + \frac{1}{2} \cdot \Gamma \cdot \Delta OASF^2$$

We can also estimate the price impact of parallel shifts and (steepening or flattening) twists in the interest rate curve using its duration and gamma. Assuming a linear change in the forward rates as a function of the term to maturity:

[9] $$\Delta r(t) = \Delta r_{shift} + t \cdot \Delta r_{twist}$$

The price impact can be estimated to the second order in shift and the first order in twist:

[10] $$\frac{P(\Delta r_{shift}, \Delta r_{twist})}{P_0} \approx 1 - D \cdot \Delta r_{shift} + \frac{1}{2} \cdot \Gamma \cdot \Delta r_{shift}^2 - \frac{1}{2} \cdot \Gamma \cdot \Delta r_{twist}$$

The twist duration is equal to half of the shift convexity. Since the shifts and twists explain the majority of the interest rate variability over short horizons, and since our modified definitions for duration and gamma are robust with respect to credit risk levels, this approximation can prove to be very useful across a wide range of market conditions.

In the rest of this paper we will focus on the interest rate dependence and denote these duration and convexity measures as $D_r$ and $\Gamma_r$, respectively.

### Credit Risk Sensitivity

The duration and convexity measures introduced in the previous subsection deal with bond price sensitivity to factors other than credit risk. The credit risk sensitivity is encoded in the dependence of the bond price on the changes in the hazard rate and recovery rate.

It is convenient to work with the continuous-time approximation to the price of a credit bond defined in Appendix A. Appendix B contains the calculated sensitivities of this price to changes in (instantaneous forward) interest rates and hazard rates which we need for our duration measures.

Before we proceed, let us introduce a risk metric borrowed from the CDS market, called “risky PV01”. This measure is instrumental for mark-to-market valuation of default swaps (see O’Kane and Turnbull [2003]), and in the continuous-time approximation it is given by:

[11] $$RPV01(T) = \int_0^T du \cdot e^{-\int_0^u (r(s)+h(s)) \cdot ds}$$

Let us define the hazard rate duration similarly to the interest rate duration, i.e. as the (minus) log-derivative of the bond price with respect to a parallel shift of its hazard rate. Comparing equations [11], [29] and [31], we obtain the following expression for the bond's hazard rate duration in terms of interest rate duration and RPV01:

[12] $$D_h = -\frac{1}{P}\cdot\frac{\partial}{\partial h}P = D_r - \left(R_p - (1-R_c)\cdot\frac{C_f}{2f}\right)\cdot P\cdot RPV01 \approx D_r\cdot(1-R_p\cdot P)$$

where the last "ballpark" approximation works quite well in most cases.

In the survival-based valuation framework, the hazard rate duration replaces the commonly used "spread duration" as a measure of credit sensitivity. Because conventionally defined spreads are a biased measure of credit risk, as demonstrated by eq. [6], then measuring bond's price sensitivity to changes in spread would also suffer from the same biases.

One could, in principle, define an unbiased credit sensitivity measure different from hazard rate duration – but it would have to rely on unbiased measures of spread such as the par spread or bond-implied CDS spread (BCDS). The latter measure, introduced in the Part 1 of this series, is the best candidate because BCDS term structure is tied directly to the term structure of hazard rates.

Using the continuous-time approximation for the BCDS spread (eq. [] in Appendix A), we can estimate the BCDS duration, which measures the relative price change of a credit bond per unit change in BCDS under the condition that both the bond price change and the BCDS change occur due to the same change of the underlying hazard rate. Considering the sensitivity of BCDS to hazard rates (eq. [] in Appendix B) and noting that the dependence of BCDS on interest rates is of lower order of magnitude compared to its dependence on hazard rates, we obtain the following expression for the BCDS duration:

[13] $$D_{BCDS} = \left(\frac{\partial BCDS}{\partial h}\right)^{-1}\cdot\frac{1}{P}\cdot\frac{\partial}{\partial h}P \approx \frac{1-R_p\cdot P}{1-R_p}D_r$$

Note that for bonds trading at premium price the BCDS duration is less than the interest rate duration, while for bonds trading at a discount it is greater than the interest rate duration. While BCDS duration has some features resembling the conventional spread duration and may thus provide for continuity in investor's intuition, we believe that for risk management purposes it is safer to use the hazard rate duration which is unencumbered with additional assumptions that were made in deriving eq. [13].

Next, let us calculate the recovery duration defined as the fractional price sensitivity to changes in the projected recovery rate (but without a re-calibration of the hazard rates), where the "ballpark" approximation works best for distressed bonds:

[14] $$D_R = \frac{1}{P}\cdot\frac{\partial}{\partial R_p}P = \frac{1}{(1-R_p)\cdot P}\cdot RPV01\cdot BCDS \approx \frac{1-P}{(1-R_p)\cdot P}$$

Finally, borrowing from CDS nomenclature again, we define the value-on-default (VOD) risk of a bond which measures the percent loss in case of instantaneous default:

[15] $$VOD = 1-\frac{R_p}{P}$$

Since RPV01 is manifestly positive (and is usually of the same order of magnitude as the interest rate duration), we conclude that the hazard rate duration is shorter than the interest rate duration. This means that distressed bonds become insensitive not only to interest rates, as demonstrated in the previous subsection, but also to hazard rates. Indeed, once the market prices them to "imminent default", any further increase in hazard rate bears no additional loss for the bond's price. Since the bond price in this scenario is already close to the expected recovery rate, the VOD risk is also very small.

On the other hand, the recovery sensitivity of the bond grows with distress levels proportionally to BCDS. For low levels of credit risk, when BCDS is typically of the order of a few tens or hundreds of basis points, the recovery sensitivity is very small. But for high levels of credit risk BCDS can grow as high as tens of percent, making the recovery rate sensitivity a sizeable number. This should not be surprising since in this case the same change of recovery value is being compared with a much lower initial price level.

Figure 2 shows the dependence of the interest, hazard and recovery rate durations, VOD and RPV01 upon the level of credit risk. We consider a hypothetical 5-year bond with 5% coupon, assuming a flat 4% LIBOR discount rate, flat term structure of hazard rates and 40% recovery rate. The BCDS spread is plotted along the bottom x-axis, and the corresponding hazard rate is plotted along the top x-axis. All durations are plotted against the left y-axis, and the bond price and VOD are plotted against the right y-axis.

Let us now turn to the convexity measures. In Appendix B we have derived the second-order derivatives of the bond price with respect to interest rates and hazard rates. Comparing equations [30] and [32] with each other and with the definition of the RiskyPV01 in [11], we obtain the following relationship between the interest and hazard rate convexities:

[16] $$\Gamma_h = \Gamma_r + 2 \cdot \left( R_p - (1 - R_c) \cdot \frac{C_f}{2f} \right) \cdot \frac{1}{P} \cdot \frac{\partial}{\partial h} RPV01$$

Since RPV01 is a decreasing function of the hazard rate, we conclude that the hazard rate convexity is lower than the interest rate convexity for cash bonds.

**Figure 2. Interest, hazard and recovery rate durations, RPV01 and price as functions of hazard rate and BCDS spread**

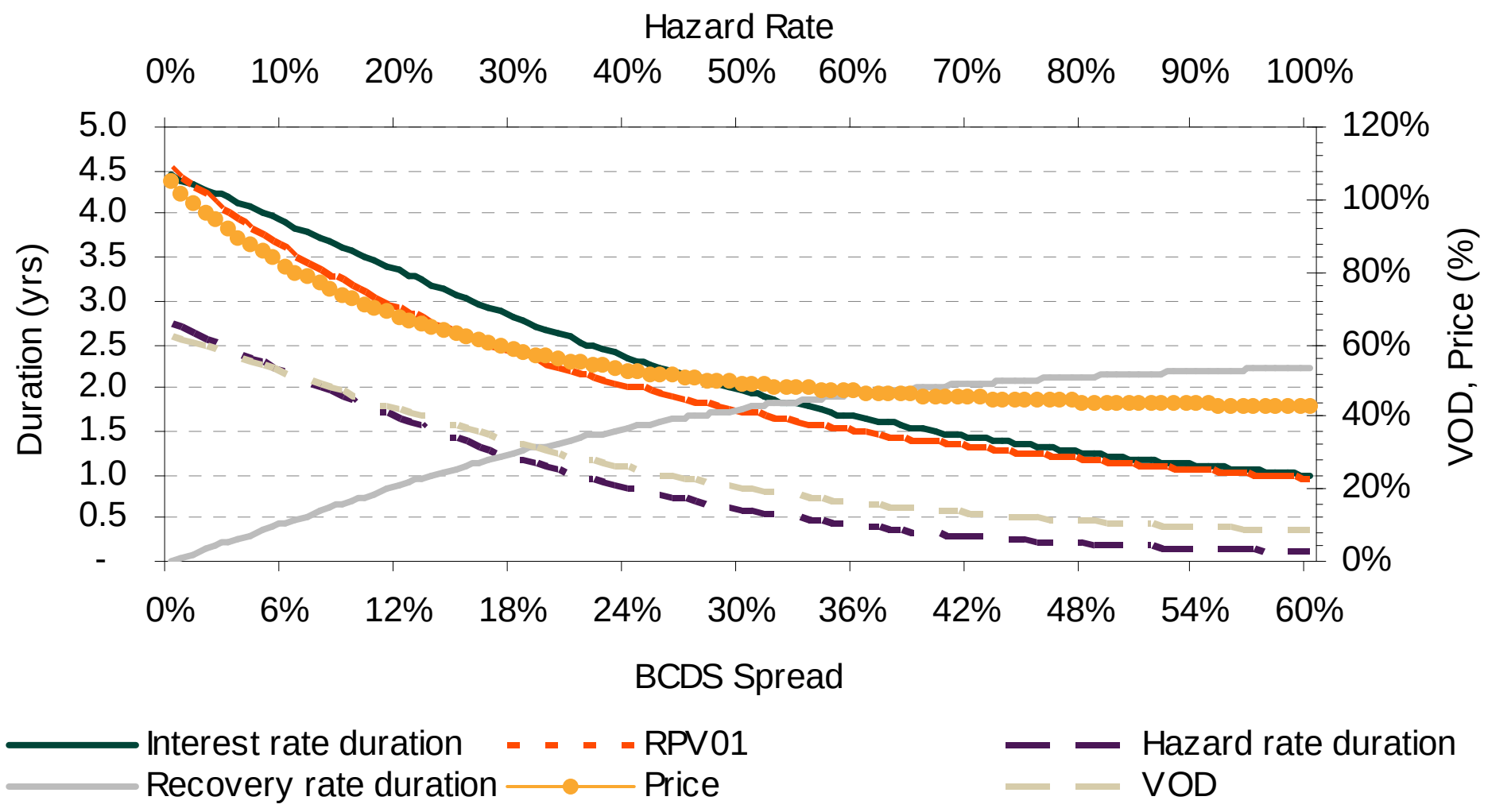

## CONSTRUCTING WELL-HEDGED BOND PORTFOLIOS

Let us now turn to the last question posed in the Introduction. How does one monetize a relative value view? First, in order to avoid inadvertently producing positive or negative returns due to coincidental market timing, one must formulate the strategy in a market-neutral way, using long-short trades. This is necessary while back-testing – in the actual investment process there could be a mixture of directional views (i.e. tactical asset allocation) together with relative value views (i.e. sector and security selection).

Secondly, when defining the long-short relative value trades one must first determine which risks need to be hedged, or in other words with respect to which market factors do we wish to be "market neutral" and for what time horizon. It is often impossible (or impractical) to be hedged with respect to all market factors. The choice of the hedge will be determined by the expected holding horizon of the relative value trade.

For example, if one expects a fast convergence to fair value then the correct hedge is with respect to sensitivity to the most volatile market factors during the short term – interest rates and spreads. If, however, the trade is not expected to converge fast and could be held to maturity in order to realize the perceived relative value, then the correct hedge is with respect to the long-run risk factors, including the idiosyncratic credit event risk and recovery.

Before we can set up a market-neutral trade, i.e. a trade that is well hedged with respect to various risk factors except the one for which we have a rich/cheap signal, we must first clarify the rules for aggregating the sensitivities for a portfolio of bonds. It can be proved that the usual aggregation rules apply and need not be modified when using the survival-based risk measures instead of the conventional ones – the portfolio duration and convexity with respect to a market risk factor are equal to the market value weighted average of constituent security durations and convexities, respectively.

Indeed, the portfolio market value is equal to a sum of market values of constituent bonds, which in turn are just their quantities $q_i$ times their prices $P_i$. While we discard the strippable cash flow valuation methodology when it comes to a single credit-risky bond, the sum of market values rule still applies to a portfolio of securities. Note that we omit the accrued interest from all calculations because its value is insensitive to all risks (except the risk of immediate default) and may be excluded from duration and convexity calculations.

[17] $$MV_{port} = \sum_{i=1}^{N} q_i \cdot P_i$$

The portfolio duration and convexity with respect to a market risk factor F are defined in the same manner as those for a single bond, which leads to the aggregation rule stated above (the sign $\kappa$ is equal to -1 for interest and hazard rate durations, and +1 for recovery duration):

[18] $$D_{port} = \frac{\kappa}{MV_{port}} \cdot \frac{\partial MV_{port}}{\partial F} = \sum_{i=1}^{N} \frac{q_i \cdot P_i}{MV_{port}} \cdot \frac{\kappa}{P_i} \cdot \frac{\partial P_i}{\partial F} = \sum_{i=1}^{N} w_i \cdot D_i$$

[19] $$\Gamma_{port} = \frac{\kappa}{MV_{port}} \cdot \frac{\partial^2 MV_{port}}{\partial F^2} = \sum_{i=1}^{N} \frac{q_i \cdot P_i}{MV_{port}} \cdot \frac{\kappa}{P_i} \cdot \frac{\partial^2 P_i}{\partial F^2} = \sum_{i=1}^{N} w_i \cdot \Gamma_i$$

Finally, the VOD risk of the portfolio is also equal to the market value weighted sum of VOD risks of the individual bonds. The aggregated VOD risk has a meaning of a percentage loss in case of simultaneous default of all bonds in the portfolio and as such it only makes sense for a portfolio consisting of bonds of the same issuer (or perhaps of several issuers all of which are driven to default by the same exogenous factor).

[20] $$VOD_{port} = 1 - \frac{\sum_{i=1}^{N} q_i \cdot R_i}{MV_{port}} = \sum_{i=1}^{N} \frac{q_i \cdot P_i}{MV_{port}} \cdot \left(1 - \frac{R_i}{P_i}\right) = \sum_{i=1}^{N} w_i \cdot VOD_i$$

Let us now derive a generic recipe for setting up market-neutral trades using credit risky bonds. The market-neutral trade is a zero-cost long-short portfolio for which the portfolio durations with respect to the selected set of risk factors are equal to zero.

Generally speaking, such a trade would not be possible to construct unless one also allows some amount of (long or short) cash position in the portfolio. Thus, the portfolio which we consider will contain N bonds plus a cash position. Let us also choose a target set of K risk factors with respect to which we wish to immunize the portfolio. These could be, for example, the interest rate and hazard rate durations, twist sensitivities, VOD, etc. We will assume that all the chosen target risk factors satisfy the market value weighted aggregation rules. Denoting the bond weights as $w_i$, and the k-th risk sensitivity of the i-th bond as $\delta_i^k$, we can rewrite the market-neutral trade definition in the following fashion:

[21] $$\begin{pmatrix} w_0 \\ w_1 \\ \vdots \\ w_N \end{pmatrix} = \begin{pmatrix} -\left(\sum_{i=1}^{N} \nu_i\right) \Big/ W \\ \nu_1 / W \\ \vdots \\ \nu_N / W \end{pmatrix}, \qquad \text{where} \quad W = \max\left(\sum_{i=1}^{N} \nu_i^+, \sum_{i=1}^{N} \left(-\nu_i^-\right)\right)$$

Here, we introduced the auxiliary variables $\nu_i$ which have the meaning of non-normalized weights of the bond positions, and $\nu_i^+$ and $\nu_i^-$ stand for the positive and negative values, respectively. The normalization factor $W$ is defined so that the long-only and short-only sub-portfolios each have a total weight of 1 (i.e. the trade is not levered), and the cash weight $w_0$ is chosen to explicitly solve the zero-cost constraint. The auxiliary variables satisfy the following equation, which guarantees that the portfolio is market-neutral:

[22] $$\begin{pmatrix} \delta_1^1 & \cdots & \delta_N^1 \\ \vdots & \ddots & \vdots \\ \delta_1^K & \cdots & \delta_N^K \\ 1 & \cdots & 1 \end{pmatrix} \cdot \begin{pmatrix} \nu_1 \\ \vdots \\ \nu_N \end{pmatrix} = \begin{pmatrix} 0 \\ \vdots \\ 0 \\ 1 \end{pmatrix}$$

Equation [22] does not have a unique solution unless $N = K + 1$ and the matrix of sensitivities on the left hand side is not degenerate. However, we can approach the problem in a slightly more general fashion by treating equation [22] not as an exact condition but rather as an approximate equation akin to a regression, where $N \le K + 1$, and the auxiliary variables $\nu_i$ are chosen so that they minimize the deviations from each of the target values on the right hand side. It is useful in this case to take $\delta_i^k$ to be "normalized" risk sensitivities, where the normalizing denominator is equal to the target accuracy desired for each risk factor. For example, if we choose the first set of sensitivities $\delta_i^1$ equal to the interest rate durations divided by 0.1 this would correspond to an implicit target accuracy of the duration-neutral solution of approximately 0.1 years of duration. With this in mind, we can write down the solution in a matrix form, denoting the sensitivity matrix as $\Lambda$:

[23] $$\begin{pmatrix} \nu_1 \\ \vdots \\ \nu_N \end{pmatrix} = \left(\Lambda' \cdot \Lambda\right)^{-1} \cdot \begin{pmatrix} 1 \\ \vdots \\ 1 \end{pmatrix}$$

Equations [21] and [23] define the optimal zero-cost long-short portfolio which is immunized with respect to a set of chosen market risk factors. If the exact solution is possible, it will be attained by this formula, but even if it is not possible, the formula still remains useful.

How big a difference does the new definition of durations make for setting up market-neutral trades? Consider for example the portfolio shown in Figure 3. We have selected three bonds issued by Kraft Foods, and constructed barbell trades using the survival-based methodology and the conventional, spread-based methodology.

For the survival-based methodology, we have immunized the trade with respect to interest rate shifts by targeting zero interest rate duration, with respect to common shifts of the sector-specific hazard rate curve by targeting zero hazard rate duration, and with respect to large-scale issuer credit deterioration by targeting zero VOD. The resulting optimal trade nets 2.56 points upfront in cash plus 30 bp running in carry.

For the spread-based methodology, we have targeted zero modified adjusted duration and also balanced the total price. The resulting barbell portfolio weights are different from those obtained in the survival-based methodology, leading to different net relative value assessment (similar carry but no upfront points – a difference of $2.56 of instantaneous profit realization for a $100 trade notional). This is despite the fact that the bonds under consideration had only moderate price premiums, and the conventional Z-Spreads were close to survival-based par LIBOR spreads (P-Spreads). Had we considered a case of high yield bonds, the discrepancy between the two approaches would have been even greater.

**Figure 3. Barbell trades with KFT bonds as of 6/30/2004**

| | | | *Survival-Based Methodology* | | | | | *Spread-Based Methodology* | | |
|---|---|---|---|---|---|---|---|---|---|---|
| **Description** | **Maturity (yrs)** | **Price** | **P-Spread (bp)** | **Interest Rate Dur.** | **Hazard Rate Dur.** | **VOD** | **Portfolio MV %** | **Z-Spread (bp)** | **Mod. Adj. Dur.** | **Portfolio MV %** |
| KFT 4.625 11/01/2006 | 2.34 | 102.90 | 4 | 2.22 | 1.32 | 0.61 | -59.23% | -1 | 2.23 | -60.04% |
| KFT 6.25 6/01/2012 | 7.92 | 104.90 | 66 | 6.24 | 3.60 | 0.62 | 97.44% | 64 | 6.34 | 100.00% |
| KFT 6.50 11/01/2031 | 27.35 | 100.44 | 79 | 11.64 | 6.76 | 0.60 | -40.77% | 80 | 12.53 | -39.96% |
| Cash | 0 | 100.00 | 0 | 0 | 0 | 0 | 2.56% | 0 | 0 | 0% |
| **Total: Portfolio** | | | 30 | 0.02 | -0.03 | 0.00 | 0% | 33 | 0.00 | 0% |

## CONCLUSIONS

In this paper we redefined and substantially expanded the set of risk and sensitivity measures for credit bonds. In particular, the new definitions of bond duration with respect to interest rates and hazard rate replace the conventional modified duration and OAS duration measures. We demonstrated that the difference between these measures and the conventional ones can be substantial, resulting in materially different portfolio weights and net relative value estimates when setting up long-short bond trades.

*Acknowledgements:* We would like to thank Marco Naldi as well as many other colleagues at Lehman Brothers Fixed Income Research department for numerous helpful discussions throughout the development and implementation of the survival-based methodology during the past several years.

## APPENDIX A: CONTINUOUS-TIME APPROXIMATION FOR CREDIT BOND PRICING

Continuous compounding is a convenient technique which may often simplify the analysis of relative value and forward pricing of credit bonds. It corresponds to coupon payments being made continuously. Consequently, in case of default such a bond would only lose an infinitesimal portion of its interest payments.

The present value for a hypothetical continuously compounded credit-risky bond can be calculated using the instantaneous forward interest rates $r(t)$ and hazard rates $h(t)$.

The base discount function is given by:

[24] $$Z(t) = \exp\left( -\int_0^t r(s) \cdot ds \right)$$

Under the Poisson model of exogenous default which is common to all reduced-form models, the survival probability is related to default hazard rate by a similar relationship:

[25] $$Q(t) = \exp\left( -\int_0^t h(s) \cdot ds \right)$$

The hazard rate, $h(t)$, can in general be stochastic, just as the spreads and the interest rates can. However, throughout we work with "breakeven" hazard rates which are deterministic (while the default event is still random).

Assuming uncorrelated interest, hazard and recovery rates, one can combine equations [24] and [25] to obtain a continuously compounded analog of the bond pricing equation [2]:

[26] $$P(T) = C \cdot \int_0^T du \cdot e^{-\int_0^u (r(s)+h(s)) \cdot ds} + e^{-\int_0^T (r(s)+h(s)) \cdot ds} + R_p \cdot \int_0^T du \cdot h(u) \cdot e^{-\int_0^u (r(s)+h(s)) \cdot ds}$$

This simple formula overestimates the present value of a credit bond for two distinct reasons:

- First, it neglects the expected coupon loss in case of default.
- Second, it overestimates the present value of the regular coupon payments because it presumes that portions of the coupon were paid earlier and it discounts those portions with a correspondingly smaller discount factor (and higher survival probability).

The more accurate approximation which we derive here corrects for these two biases.

The correction for the coupon loss bias can be estimated by noting that the expected timing of the default event under a constant hazard rate assumption is roughly half-way through the payment period. Hence, the expected loss of the accrued interest equals to approximately half of the scheduled coupon payment, which in turn is equal to *1/f* fraction of the coupon rate for a bond with frequency *f*. Consequently, to correct for this bias we should explicitly subtract the expected accrued interest loss (assuming a given coupon recovery fraction $R_c$) from the principal recovery rate $R_p$ in the original formula.

The correction for the early discount bias can be estimated by noting that by distributing the coupon payments evenly between the two coupon dates we get the survival-weighted present value which is roughly half-way between the present value of a bullet coupon payment on the two ends of the coupon period. Thus, for each bullet coupon the continuous-time formula [26] corresponds to a present value bias equal to half of difference between the "true" present value of the earlier coupon payment and the current coupon payment. When summing up all of these biases, the corrections for all intermediate coupon payments cancel each other, and the total present value bias is simply half of the difference between the present value of the first coupon payment and the last coupon payment. For the valuation date just prior to a coupon payment, this results in a simple estimate since the present value of that impending coupon payment is simply equal to its amount. For other valuation dates the situation is slightly more complicated but the approximation remains pretty close nevertheless.[1]

We obtain the continuous-time approximation for the clean price of an *f*-frequency credit bond by subtracting these two bias estimates from the original "naïve" formula.

Finally, we should also include the OAS-to-Fit (OASF), an issue-specific discounting measure introduced in Part 1 of this series that allows us to use the issuer- or sector-specific hazard rate term structure while exactly fitting the observed price of individual bonds.

The final formula for the clean price of a fixed-coupon credit bond in the continuous-time approximation is:

[27]
$$\begin{aligned} P(T|f) \approx\ & C_f \cdot \int_0^T du \cdot e^{-\int_0^u (r(s)+h(s)+OASF)\cdot ds} + e^{-\int_0^T (r(s)+h(s)+OASF)\cdot ds} \\ & - \frac{C_f}{2f}\cdot\left(1 - e^{-\int_0^T (r(s)+h(s)+OASF)\cdot ds}\right) \\ & + \left(R_p - (1-R_c)\cdot\frac{C_f}{2f}\right)\cdot\int_0^T du\cdot h(u)\cdot e^{-\int_0^u (r(s)+h(s)+OASF)\cdot ds} \end{aligned}$$

This approximation is quite accurate across all values of coupons and for all shapes and levels of the underlying interest rate and hazard rate curves. Both correction terms can be quite important. The coupon loss bias term becomes zero when the coupon recovery is equal to 1. However, in practice we often assume the coupon recovery rate equal to zero and therefore this correction is not negligible.

For completeness, let us also write down the continuous-time approximation to the bond-implied CDS spread (BCDS) which was introduced in the Part 1 of this series and will be a subject of a detailed investigation in Part 3. We consider BCDS to be the best un-biased spread measure for a credit bond that should replace the conventional LIBOR spread:

[28]
$$BCDS(T) = (1-R_p)\cdot\frac{\int_0^T du\cdot h(u)\cdot e^{-\int_0^u (r(s)+h(s))\cdot ds}}{\int_0^T du\cdot e^{-\int_0^u (r(s)+h(s))\cdot ds}} + OASF$$

---

[1] *For valuation dates that fall between coupon payment dates, the approximate present value of the first coupon payment is non-trivial because the conventional definition of the "clean" price depends on the linear coupon accrual, while the correct present value calculation involves a discounting which is closer to an exponential formula. We ignore this additional discrepancy in our approximation.*

## APPENDIX B: CONTINUOUS-TIME APPROXIMATION FOR SENSITIVITIES

Here we use the continuous-time approximation to the price of the bond defined in Appendix A to derive formulas for first- and second-order price sensitivities to interest rates, hazard rates, and recovery rates. These formulas will allow us to uncover useful relationships between the various "durations" and "convexities" of credit bonds.

[29]
$$\begin{aligned}\frac{\partial}{\partial r}P(T|f) = & -C_f \cdot \int_0^T du \cdot u \cdot e^{-\int_0^u (r(s)+h(s)+OASF)\cdot ds} - T \cdot \left(1+\frac{C_f}{2f}\right)\cdot e^{-\int_0^T (r(s)+h(s)+OASF)\cdot ds} \\ & - \left(R_p - (1-R_c)\cdot\frac{C_f}{2f}\right)\cdot \int_0^T du \cdot u \cdot h(u)\cdot e^{-\int_0^u (r(s)+h(s)+OASF)\cdot ds}\end{aligned}$$

[30]
$$\begin{aligned}\frac{\partial^2}{\partial r^2}P(T|f) = & \; C_f \cdot \int_0^T du \cdot u^2 \cdot e^{-\int_0^u (r(s)+h(s)+OASF)\cdot ds} + T^2 \cdot \left(1+\frac{C_f}{2f}\right)\cdot e^{-\int_0^T (r(s)+h(s)+OASF)\cdot ds} \\ & + \left(R_p - (1-R_c)\cdot\frac{C_f}{2f}\right)\cdot \int_0^T du \cdot u^2 \cdot h(u)\cdot e^{-\int_0^u (r(s)+h(s)+OASF)\cdot ds}\end{aligned}$$

[31]
$$\begin{aligned}\frac{\partial}{\partial h}P(T|f) = & -C_f \cdot \int_0^T du \cdot u \cdot e^{-\int_0^u (r(s)+h(s)+OASF)\cdot ds} - T \cdot \left(1+\frac{C_f}{2f}\right)\cdot e^{-\int_0^T (r(s)+h(s)+OASF)\cdot ds} \\ & - \left(R_p - (1-R_c)\cdot\frac{C_f}{2f}\right)\cdot \int_0^T du \cdot u \cdot h(u)\cdot e^{-\int_0^u (r(s)+h(s)+OASF)\cdot ds} \\ & + \left(R_p - (1-R_c)\cdot\frac{C_f}{2f}\right)\cdot \int_0^T du \cdot e^{-\int_0^u (r(s)+h(s)+OASF)\cdot ds}\end{aligned}$$

[32]
$$\begin{aligned}\frac{\partial^2}{\partial h^2}P(T|f) = & \; C_f \cdot \int_0^T du \cdot u^2 \cdot e^{-\int_0^u (r(s)+h(s)+OASF)\cdot ds} + T^2 \cdot \left(1+\frac{C_f}{2f}\right)\cdot e^{-\int_0^T (r(s)+h(s)+OASF)\cdot ds} \\ & + \left(R_p - (1-R_c)\cdot\frac{C_f}{2f}\right)\cdot \int_0^T du \cdot u^2 \cdot h(u)\cdot e^{-\int_0^u (r(s)+h(s)+OASF)\cdot ds} \\ & - 2\cdot\left(R_p - (1-R_c)\cdot\frac{C_f}{2f}\right)\cdot \int_0^T du \cdot u \cdot e^{-\int_0^u (r(s)+h(s)+OASF)\cdot ds}\end{aligned}$$

[33]
$$\frac{\partial}{\partial R_p}P(T|f) = \int_0^T du \cdot h(u)\cdot e^{-\int_0^u (r(s)+h(s)+OASF)\cdot ds}$$

[34]
$$\frac{\partial}{\partial OASF}P(T|f) = \frac{\partial}{\partial r}P(T|f)$$

[35]
$$\frac{\partial}{\partial h}BCDS(T) = 1 - R_p + \frac{\partial}{\partial r}BCDS(T)$$

## APPENDIX C: THE CORRELATION BETWEEN RATES AND SPREADS

Here we briefly recap some of the results from Berd and Silva (2004) regarding the long-term correlation between the (conventionally defined) credit spreads and interest rates. Figure A-1 shows the time series of the Lehman Brothers Credit Index OAS vs. the Treasury curve shift and twist factors (as defined below). We show the incremental changes in shift and steepness (negative twist) from Jan 1990. The shift and twist factors of the Treasury curve are defined using the 2, 5, 10, 20 and 30-year benchmark Treasury yields.

[36]
$$\begin{cases} \Delta F_{shift} = \frac{1}{5}\cdot\left(\Delta y_2 + \Delta y_5 + \Delta y_{10} + \Delta y_{20} + \Delta y_{30}\right) \\ \Delta F_{twist} = \frac{1}{10}\cdot\left(2\cdot\Delta y_2 + \Delta y_5 - \Delta y_{20} - 2\cdot\Delta y_{30}\right) \end{cases}$$

To quantify the correlation of interest rates and credit spreads, we used the Lehman Brothers multi-factor credit risk model (Naldi *et. al.* [2002]). This is an econometric model that decomposes corporate bond returns into a linear combination of a number of factors including six Treasury (key-rate) factors, six swap spread factors, and a number of credit spread factors. The spread factors include 27 industry/rating sector factors measured in terms of LIBOR OAS, a spread twist factor which captures spread curve steepening or flattening, and an OAS dispersion factor which captures the dependence of bond's returns on its relative OAS to the sector average. The model estimates the covariance matrix of all common driving factors, as well as the issuer-specific risk of bonds belonging to each industry/rating sector. Our analysis was based on the model estimates as of the end of April 2003.

In order to take into account the issuer-specific risk and an incomplete diversification of typical investors' portfolios, we defined a sector portfolio to consist of 20 equally weighted bonds having on average the same maturity and same OAS as the corresponding sector. By construction of the risk model, such portfolio is not exposed to spread twist or OAS dispersion factors. The estimates of the correlations between the OAS changes of these hypothetical sector portfolios with the Treasury shift and twist factors are shown in Figures A-2, A-3, where we show two sets of numbers, one for the correlations of the Treasury spreads with Treasury curve factors, and the other for the correlation of LIBOR spreads with swap curve factors. As we can see, all correlations are negative and quite significant. Their magnitude is greater than the "optical" correlation effect discussed in this paper can explain, suggesting that there is also a fundamental negative correlation between interest rates and credit risk, perhaps driven by common macro-economic factors affecting both markets.

**Figure A-1. Lehman Credit Index OAS vs. Treasury curve shift and twist**

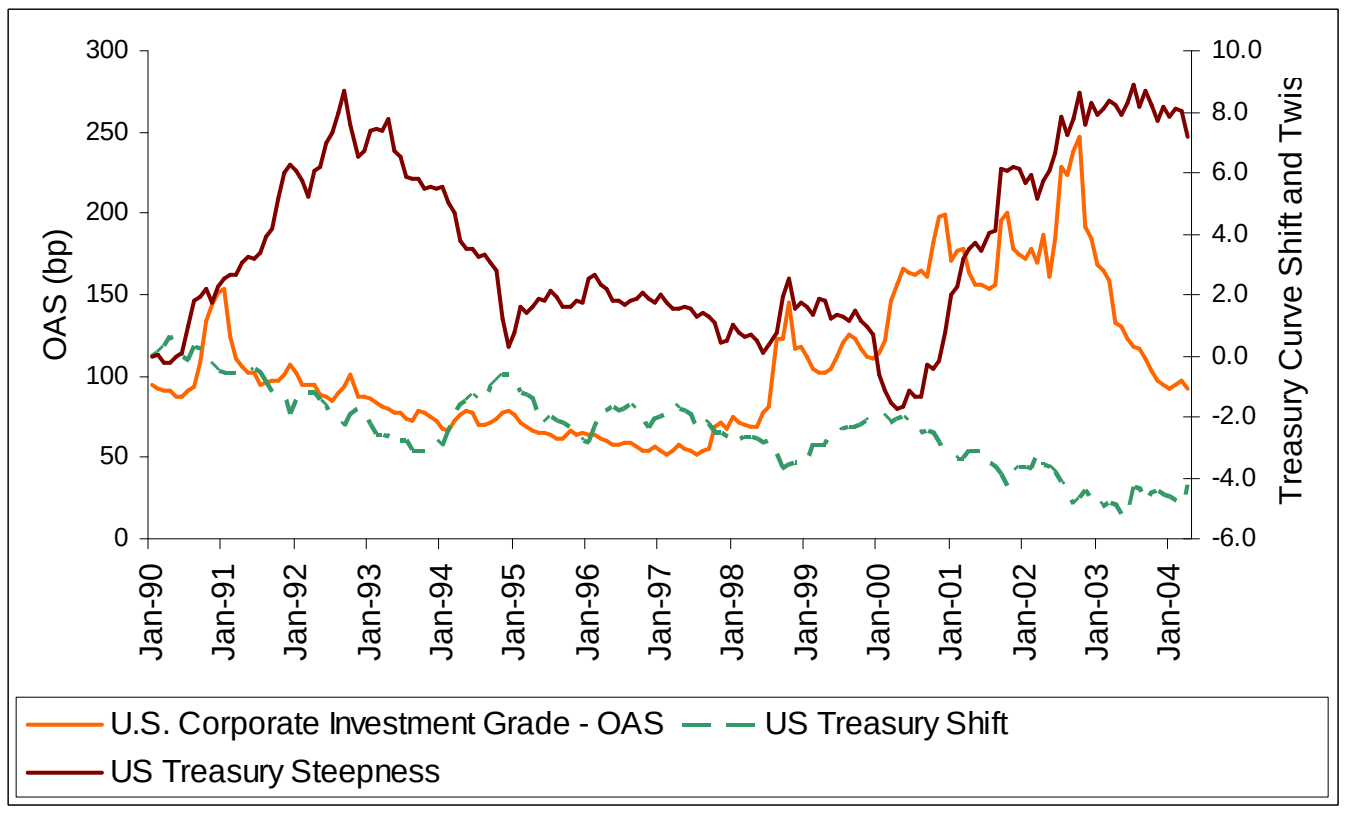

**Figure A-2a: Industry Portfolio Spread Correlations with Treasury Curve Shifts**

| | AAA/AA | A | BBB |
|---|---|---|---|
| FINANCIALS | | | |
| Banking and Brokerage | -18% | -24% | -20% |
| Financial Companies, Insurance and REITS | -23% | -28% | -26% |
| INDUSTRIALS | | | |
| Basic Industries and Capital Goods | -16% | -22% | -24% |
| Consumer Cyclicals | -14% | -27% | -23% |
| Consumer Non-Cyclicals | -17% | -19% | -21% |
| Communication and Technology | -17% | -21% | -24% |
| Energy and Transportation | -17% | -23% | -26% |
| UTILITIES | -17% | -20% | -20% |
| NON-CORPORATE | -4% | -17% | -21% |

**Figure A-3a: Industry Portfolio Spread Correlations with Treasury Curve Twists (Flattening)**

| | AAA/AA | A | BBB |
|---|---|---|---|
| FINANCIALS | | | |
| Banking and Brokerage | -16% | -19% | -16% |
| Financial Companies, Insurance and REITS | -19% | -24% | -22% |
| INDUSTRIALS | | | |
| Basic Industries and Capital Goods | -8% | -17% | -19% |
| Consumer Cyclicals | -9% | -23% | -20% |
| Consumer Non-Cyclicals | -11% | -12% | -16% |
| Communication and Technology | -12% | -16% | -20% |
| Energy and Transportation | -14% | -16% | -20% |
| UTILITIES | -9% | -15% | -16% |
| NON-CORPORATE | -3% | -11% | -17% |

**Figure A-2b: Industry Portfolio L-OAS Correlations with Swap Curve Shifts**

| | AAA/AA | A | BBB |
|---|---|---|---|
| FINANCIALS | | | |
| Banking and Brokerage | -36% | -38% | -25% |
| Financial Companies, Insurance and REITS | -42% | -40% | -36% |
| INDUSTRIALS | | | |
| Basic Industries and Capital Goods | -37% | -41% | -37% |
| Consumer Cyclicals | -36% | -41% | -29% |
| Consumer Non-Cyclicals | -38% | -38% | -35% |
| Communication and Technology | -33% | -35% | -32% |
| Energy and Transportation | -37% | -41% | -39% |
| UTILITIES | -33% | -36% | -28% |
| NON-CORPORATE | -31% | -33% | -28% |

**Figure A-3b: Industry Portfolio L-OAS Correlations with Swap Curve Twists (Flattening)**

| | AAA/AA | A | BBB |
|---|---|---|---|
| FINANCIALS | | | |
| Banking and Brokerage | -29% | -32% | -21% |
| Financial Companies, Insurance and REITS | -35% | -34% | -30% |
| INDUSTRIALS | | | |
| Basic Industries and Capital Goods | -28% | -31% | -29% |
| Consumer Cyclicals | -29% | -32% | -24% |
| Consumer Non-Cyclicals | -28% | -29% | -27% |
| Communication and Technology | -25% | -27% | -25% |
| Energy and Transportation | -30% | -31% | -31% |
| UTILITIES | -24% | -28% | -23% |
| NON-CORPORATE | -22% | -27% | -22% |